\documentclass{aastex62}
\usepackage{savesym}
\usepackage{amsfonts,amsmath,graphicx,natbib,subfigure,color,longtable,units}

\savesymbol{tablenum}
\usepackage{siunitx}
\restoresymbol{SIX}{tablenum}
\DeclareSIUnit\parsec{pc}
\DeclareSIUnit\year{yr}
\DeclareSIUnit\mag{mag}

\submitjournal{ApJ}
\received{1/24/19}
\revised{3/8/19}
\accepted{3/10/19}
\shorttitle{The Local Perspective on the Hubble Tension}
\shortauthors{Kenworthy et al.}

\begin{document}

\title{The Local Perspective on the Hubble Tension: Local Structure Does Not Impact Measurement of the Hubble Constant}

\correspondingauthor{D'Arcy Kenworthy}
\email{wkenwor1@jhu.edu, darcy@darcykenworthy.com}

\author{W. D'Arcy Kenworthy}
\affil{Department of Physics and Astronomy, Johns Hopkins University \\
3701 San Martin Drive \\
Baltimore, MD 21218, USA}
\noaffiliation

\author{Dan Scolnic}
\affil{Department of Physics, Duke University\\
120 Science Drive\\
Durham, NC, 27708, USA}

\author{Adam Riess}
\affil{Department of Physics and Astronomy, Johns Hopkins University \\
3701 San Martin Drive \\
Baltimore, MD 21218, USA}
\affil{Space Telescope Science Institute \\
3700 San Martin Drive\\
Baltimore, MD 21218, USA}



\begin{abstract}

We use the largest sample to date of spectroscopic SN Ia distances and redshifts to look for evidence in the Hubble diagram of large scale outflows caused by local voids suggested to exist at $z<0.15$. Our sample combines data from the Pantheon sample with the Foundation survey and the most recent release of lightcurves from the Carnegie Supernova Project to create a sample of 1295 SNe over a redshift range of $0.01<z<2.26$. We make use of an inhomogeneous and isotropic Lemaitre-Tolman-Bondi metric to model a void in the SN Ia distance-redshift relation. We conclude that the SN luminosity distance-redshift relation is inconsistent at the $4-5\sigma$ confidence level with large local underdensities ($|\delta| > 20\%$, where the density contrast $\delta = \Delta \rho /\rho$) proposed in some galaxy count studies, and find no evidence of a change in the Hubble constant corresponding to a void with a sharp edge in the redshift range $0.023<z<0.15$. {With empirical precision of $\sigma_{H_0} = 0.60\%$, we conclude that the distance ladder measurement is not affected by local density contrasts, in agreement with cosmic variance of $\sigma_{H_0} = 0.42\%$ predicted from simulations of large-scale structure. Given that uncertainty in the distance ladder value is $\sigma_{H_0}=2.2\%$, this does not affect the Hubble tension.}  We derive a $5\sigma$ constraint on local density contrasts on scales larger than $\SI{69}{\mega\parsec}\ h^{-1}$ of $|\delta| <  27\%$. The presence of local structure does not appear to impede the possibility of measuring the Hubble constant to 1\% precision.

\end{abstract}



\section{Introduction} \label{sec:intro}

When interpreting measurements based on type Ia supernovae (SNe Ia), care must be taken to distinguish between cosmological effects and potential cosmic variance stemming from matter density fluctuations surrounding the Milky Way \citep{Enqvist2007TheObservations,Wojtak2014CosmicSimulations}. While inhomogeneity on small scales ($<\SI{100}{\mega\parsec}$)  is unlikely to affect measurements, inhomogeneities on large scales ($>\SI{100}{\mega\parsec}$) or giant scales ($>\SI{1000}{\mega\parsec}$) could cause biases in inference of cosmological parameters from SN data. In this context we consider the tension between local measurements of the Hubble constant and predictions from the CMB. The most recent values are the \citet{Riess2018MilkyConstant} (hereafter R18) measurement ($H_0= 73.52 \pm 1.62\ \SI{}{\kilo \metre\per\second \per\mega\parsec}$) using distance ladder methods and the \citet{PlanckCollaboration2018PlanckParameters} value ($H_0= 67.4 \pm 0.5\ \SI{}{\kilo \metre\per\second \per\mega\parsec}$) found using the $\Lambda$CDM model and  based on CMB power spectrum measurements. The 8.7\% difference between these values is a disagreement of $3.6\sigma$ ($p<0.05\%$). The Hubble tension is bolstered by independent measurements of $H_0$ at different redshifts which suggest a schism between the early and late universe. Results from the H0LiCOW collaboration based on time delay distance measurements of lensed quasars at $z\leq 1$ give $H_0= 72.5^{+2.1}_{-2.3}\ \SI{}{\kilo \metre\per\second \per\mega\parsec}$ \citep{Birrer2018H0LiCOWConstant}. This cosmographic result is independent of and in agreement with the R18 result. Meanwhile, \citet{Addison2018ElucidatingDiscrepancy} combines galaxy and Ly$\alpha$ measurements of baryon acoustic oscillations calibrated through the early universe physics of Big Bang nucleosynthesis with precise estimates of the primordial deuterium abundance to produce a value of $H_0= 66.98 \pm 1.18\ \SI{}{\kilo \metre\per\second \per\mega\parsec}$ under $\Lambda$CDM, yielding a measurement independent of and consistent with the Planck+$\Lambda$CDM measurement. As several different methods have been used to measure local and early universe values of $H_0$, while remaining in tension, we require either new physics or unknown systematics in \textit{multiple} experiments to resolve the tension.


A systematic error in the local distance ladder could result if the Milky Way lies in a large region of greatly negative density contrast, defined as  $\delta=\frac{\rho(r)-\bar{\rho}}{\bar{\rho}}$ where $\bar{\rho}$ is the mean matter density at the homogeneity scale and $\rho(r)$ is the spherically averaged matter density within radius $r$. A local void with the Milky Way at the center would cause nearby galaxies to have significant positive peculiar velocities, biasing local measurements above the true value of $H_0$ \citep{Fleury2017HowDiagram}. The value of $H_0$ reported by \citet{Riess2018MilkyConstant} was based on an analysis of type Ia supernovae \citep{Riess2016ACONSTANT,Scolnic2015SUPERCAL:SUPERNOVAE} (hereafter the R16 sample), which applied corrections to account for visible density fluctuations across the sky based on 2M++ redshift maps \citep{Carrick2015CosmologicalField}. These corrections could be susceptible to large systematic uncertainties in local density field reconstructions \citep{Andersen2016CosmologyEffects}. However, \citet{Riess2016ACONSTANT} looked for evidence of outflow in their SN Ia sample and found no evidence of the effects of local voids, in agreement with expectations from \citet{Odderskov2014OnConstant} for cosmic variance to be present at the 0.3\% level, much less than statistical errors in determination of $H_0$.

Several teams have used analytic methods and simulations to measure the systematic error in distance-ladder measurements of the Hubble constant due to local density contrasts \citep{Marra2013CosmicParameter, Odderskov2014OnConstant, Wojtak2014CosmicSimulations}. Most recently, \citet{Wu2017SampleConstant} used the Dark Sky simulation and the R16 SN sample to quantify systematic error due to variation in the local density contrast and the spatial inhomogeneity of the SN sample. Their analysis finds a cosmic variance in the R16 estimate of 0.42\%, a factor of 20 less than the 8.3\% change in the R18 measurement necessary to reconcile the measurement with Planck. {They concluded that to resolve the full tension in $H_0$ would require a void with radius $\approx \SI{150}{\mega\parsec}$ and density contrasts of $\delta \approx -0.8$, and would be $\approx 20 \sigma$ discrepant with $\Lambda$CDM. } 

Evidence for local voids has been varied among studies using galaxy catalogs with infrared data to probe local densities. \cite{Carrick2015CosmologicalField} uses 2M++ redshift measurements and galaxy catalogs to reconstruct the local density field over the sky, finding no evidence of large-scale density contrasts within $z=0.04$. \citet{Whitbourn2014TheRedshifts} (hereafter WS14) uses SDSS and 6dFGS data covering $\approx 20\%$ of the sky, finds local matter density contrasts of $\delta=-0.4,-0.15,-0.05$, depending on the part of the sky investigated, extending to redshifts $z<0.05$. \citet{Keenan2013EVIDENCEDISTRIBUTION} (hereafter KBC), based on a sample drawn from $\approx 15\%$ of the sky, show a matter density contrast of $\delta \approx -0.3$ within redshifts $z<0.07$. These studies have some unquantified systematic uncertainties (see our discussion in Section \ref{sec:conclusion}). In \citet{Hoscheit2018TheModel}, the authors investigate the consistency of a model based on the density contrasts of KBC with the R16 SN sample, and examine the SN data for evidence of a sharp void. Similarly \citet{Shanks2019iGaia/iTension} applies corrections based on the density contrasts of WS14 to the Pantheon sample \citep{Scolnic2018TheSample} and measures the resulting Hubble constant. {They concluded that the effects of local density contrasts could explain the Hubble tension when combined with purported effects in Cepheid parallaxes. These conclusions were contested by \citet{Riess2018SevenArXiv:1810.02595}.}

Thus, while large scale structure simulations conclude a significant local void is exceedingly unlikely, and there is little previous evidence of large-scale outflow from the SN Hubble diagram, some teams have claimed excess structure using data covering a minority of the sky. Our work focuses on assessing evidence for outflow from and the consistency of large local void models with the best populated SN Hubble diagram to date. In section \ref{sec:snsample}, we will discuss our sample of 1295 unique cosmologically useful SNe from the Pantheon, Foundation, and Carnegie Supernova Project samples. This sample allows us to probe local structure over a range of redshifts with greater precision than previous studies. In Section \ref{sec:void}, we constrain models of a local void by assessing their consistency with this sample. In Section \ref{sec:conclusion} we discuss our results and make suggestions for further work.

\section{Combining Supernova Samples} \label{sec:snsample}

In order to analyze this problem with a larger number of supernovae at low redshifts, we create a composite sample from the Pantheon, Foundation, and CSP samples. The Pantheon sample is a successor to the earlier Supercal \citep{Scolnic2015SUPERCAL:SUPERNOVAE} sample used in R16, with 1048 SNe from the CSP, PS1, CfA1-4, SNLS, SDSS, SCP, GOODS, and CANDELS surveys in one sample with consistent photometry \citep{Scolnic2018TheSample}. Notably, in comparison with the Supercal sample used in the analysis of the Hubble flow from R16, low-z supernovae from surveys with insufficent reference stars for modern cross-calibration, such as those from the Calan/Tololo survey \citep{Hamuy1993TheSearch}, have been removed. Foundation is a survey that aims to provide a large, high-fidelity, and well calibrated sample of low-redshift cosmological SNe \citep{Foley2018TheRelease,Jones2018ShouldEnvironments}. Our sample includes 177 SN from the Foundation survey. Further, the Carnegie Supernova Project has recently released the final photometric observations for all 134 SNe observed from 2004-2009 \citep{Krisciunas2017TheExplosions}. We replace and expand upon the older CSP data incorporated into Pantheon with this newest data release.

The Foundation sample has been calibrated to the same photometric system as Pan-STARRS (Scolnic et al. in prep). To transform the CSP photometry from the natural photometric system to AB magnitudes uniformly calibrated with Pantheon, we apply Supercal corrections from \citet{Scolnic2015SUPERCAL:SUPERNOVAE} to the calibration. This procedure compares measurements of secondary standards under various photometric systems to measurements of the same stars in the Pan-STARRS1 (PS1) system, and determines offsets for each system relative to PS1. Since we have not yet repeated the Supercal analysis using the recent data release, we assume that the zero-point calibration of CSP has not significantly changed since their second data release \citep{Stritzinger2011THESUPERNOVAE}. To test this assumption, we examine the individual photometric observations of supernovae in each filter for signs of a systematic offset between Data Release 2 and Data Release 3. The largest median shift of observations is \SI{6}{\milli\mag} in \textit{i} band, corresponding to changes in median luminosity distance $<0.5\%$.

We estimate stellar masses for the CSPDR3 host galaxies for use in standardization using \textit{uvgriz} photometry from the SkyMapper and Pan-STARRS surveys \citep{Wolf2018SkyMapperDR1,Chambers2016TheSurveys} \footnote{A machine-readable table of these masses can be found at \url{https://github.com/darcykenworthy/CSP-Masses/tree/master} }.  The photometry is then fit to template spectra using the Z-PEG code from \citet{LeBorgne2002PhotometricConstraint}. More detail about the assumptions and methods can be found in \citet{Pan2014TheFactory}. Both the Pantheon and Foundation samples have already been assigned host galaxy stellar masses \citep{Jones2018ShouldEnvironments}.

The corrected peak magnitude measurements of our SNe are based on the methodology of \citet{Scolnic2018TheSample}. We use the code SNANA \citep{Kessler2009SNANA:Analysis} to fit our supernova lightcurves using the SALTII supernova light-curve model \citep{Guy2010TheConstraints, Betoule2014ImprovedSamples}, giving a flux normalization $x_0$ (converted into magnitudes as $m_B$) along with the stretch parameter $x_1$ and color parameter $c$. As in previous work  \citep{Betoule2014ImprovedSamples} we make corrections for a step in magnitude that depends on host galaxy masses, based on evidence that SN Ia are intrinsically brighter in host galaxies with mass above $M_\text{Stellar} \approx 10^{10} M_\odot$. Further, to account for expected distance biases from contamination, lightcurve-fitting, and selection bias we use BEAMS with Bias Corrections (the BBC method) \citep{Kessler2017CorrectingSamples}. For consistency with R16, in our determination of BBC bias corrections we use only the ``G10'' model for the Gaussian intrinsic scatter of SNe Ia, with 30\% of variation chromatic and 70\% achromatic  \citep{Guy2010TheConstraints}. Our final expression is then  

\begin{equation}
m_B^0=m_B+\alpha x_1-\beta c + \Delta_{M_\text{Stellar}}-\Delta_b,
\end{equation}

where $m_B^0$ is the corrected SN peak apparent magnitude, $\Delta_b$ is the bias correction derived from the BBC method, and the $\Delta_{M_\text{Stellar}}$ correction assumes an underlying function
\[ \Delta_{M_\text{Stellar}}= \gamma \begin{cases}
	\frac{1}{2} & M_\text{Stellar} > M_\text{Step} \\
	-\frac{1}{2} & M_\text{Stellar} < M_\text{Step},
\end{cases}
\]
for which we construct an estimator accounting for the uncertainty in $M_\text{Stellar}$. $M_\text{Stellar}$ is used in logarithmic units of solar mass. For consistency with R16, we choose values for our nuisance parameters $\alpha=0.14$, $\beta=3.1$, $\gamma=0.06$, $M_\text{Step}=10$. These values are consistent to within $1\sigma$ with those derived from the Pantheon sample. 
R16, \citet{Hoscheit2018TheModel}, and \citet{Shanks2019iGaia/iTension} did not budget for systematic uncertainties in the Hubble flow sample, such as uncertainties in the nuisance parameters $\alpha,\beta,\gamma,M_\text{step}$, calibration uncertainties, and  possible redshift evolution of the nuisance parameters. For a measurement of $H_0$, these systematics are expected to roughly cancel. However, a differential measurement of the Hubble constant at different redshifts is expected to be much more strongly affected by these systematic uncertainties. Further, R16 accounted for the effect of uncertainty in the deceleration parameter $q_0$ on $H_0$ by reanalyzing the data with different values of $q_0$. Since in this work we extend the redshift range of our primary fit (as discussed in \ref{subsec:interceptAnalysis}), it is appropriate to extend the R16 analysis further to budget for these uncertainties. To this end we calculate a matrix of covariances between SN corrected magnitudes $\boldsymbol{C}$ rather than simply calculating variance for each individual SN, as done in R16. 

The diagonal entries of the SN covariance matrix includes variance contributions from the flux normalization, color, stretch, host galaxy mass, peculiar velocity dispersion, spectroscopic redshift error, weak lensing uncertainty, and uncertainty from bias correction, as detailed in \citet{Kessler2017CorrectingSamples}. A final variance term is the intrinsic scatter in type Ia SNe, which we set to $\sigma_{\text{int}}=0.1$ mag, consistent with R16. The off-diagonal entries contain covariance terms based on the systematic uncertainties given in \citet{Scolnic2018TheSample}. The full Pantheon analysis, incorporating 85 separate systematic uncertainties (of which 74 relate to calibration), has only been performed on SNe from the original Pantheon set. We make the simplistic assumption that the redshift dependence of the systematics of the Pantheon sample is identical to that of the CSPDR3 and Foundation samples. We assign the covariance between new SNe as the covariance between their nearest neighbors in redshift from the Pantheon set. We also include uncertainty in the FLRW luminosity distance from an uncertainty in $q_0$ of 0.033, based on uncertainty in Pantheon measurements of $\Omega_M$ under a flat $\Lambda$CDM model (this is discussed further in Section \ref{sec:cosmo}). We also add the coherent variance terms (intrinsic SALT-II, lensing error, and peculiar velocity dispersion) to the off-diagonal elements between duplicate observations from different surveys of a single SN. For our primary fit, using 1054 SNe over a redshift range $0.01<z<0.50$, systematic uncertainties contribute $ \approx 70\%$ of the variance in our primary measurements of $\Delta a_B$ (see Section \ref{subsec:interceptAnalysis} and Table \ref{table:interceptchanges}), showing the importance of systematic uncertainties when analyzing large SN samples.

As in other modern cosmological studies \citep{Betoule2014ImprovedSamples,Riess2016ACONSTANT,Scolnic2018TheSample,Jones2018ShouldEnvironments}, to ensure the lightcurves of our SNe are well fit by the SALTII model, we make a number of cuts for lightcurve quality. These are 1) a constraint that the color parameter $-0.3<c<0.3$; 2) the lightcurve stretch parameter $-3<x_1<3$; 3) the error in stretch parameter $\sigma_{x_1}<1$; 4) the error in the rest frame peak date $\sigma_{t_0}<2$ days; 5) Milky Way reddening $E_{B-V}<0.2$; 6) a $4\sigma$-clip in Hubble residual; and 7) that the $\chi^2$ of the SALT-II fit was `good' ($\text{fitprob}>0.001$). The $\chi^2$ cut is neglected for Foundation, CSP, and SNLS, which have their own requirements for goodness-of-fit  \citep{Foley2018TheRelease,Betoule2014ImprovedSamples}. Using our selection of quality cuts, there are 1295 unique SNe in our sample, over a redshift range of $0.01<z<2.26$. We use two sets of redshift cuts in this work; over the R16 redshift range $0.023<z<0.15$ there are 398 unique SNe (as compared to 217 in R16), and over the redshift range $0.01<z<0.5$ there are 1054 unique SNe. This is the largest sample of cosmologically useful, spectroscopically confirmed SNe Ia to date.

\section{Cosmological Models} \label{sec:cosmo}

\subsection{Homogeneous cosmology} \label{sec:homogcosmo}
To examine the Hubble diagram for evidence of local structure affecting the luminosity distance-redshift relation, we require a homogeneous model to act as a baseline analysis. Any cosmology with a single homogeneous and isotropic metric requires that the metric take the FLRW form,

\begin{equation}
ds^2=c^2 dt^2 - a^2(t) (\frac{dr^2}{1-kr^2}+r^2 d\Omega^2) \label{eq:FLRWmetric},
\end{equation}
where $a(t)$ is the scale factor, $k$ is the intrinsic curvature of the metric, $\Omega$ is solid angle, and $r$ is the radial coordinate.

According to \citet{Visser2004JerkState}, using the FLRW metric, the Taylor expansion of luminosity distance as a function of redshift about $z=0$ is

\begin{equation}
D_L^{\text{FLRW}}(z;H_0,q_0,j_0)=\frac{cz}{H_0}[1+\frac{1-q_0}{2}z-\frac{1-q_0-3q_0^2+j_0-\Omega_k}{6}z^2+O(z^3)] \label{eq:redshiftexpansion},
\end{equation}

with the Hubble constant  $H_0=\dot{a}/{a}$, the cosmological deceleration parameter $q_0={\ddot{a}a}/{\dot{a}^2}$, and the cosmological jerk parameter $j_0={\dddot{a}a^2}/{\dot{a}^3}$  . We refer to this as a ``kinematic'' expansion, as it makes no inherent assumptions about the dynamics of the universe, such as the evolution of critical densities, only that the metric is homogeneous and isotropic. When using a $\Lambda$CDM model to describe the dynamical behavior, $j_0-\Omega_k=1-2\Omega_k$ while $q_0={\Omega _M}/{2}-\Omega_\Lambda$. Under flat $\Lambda$CDM, $q_0=3 \Omega_M /2 -1$ and $j_0$=1. For our baseline analysis, we follow R16 with $q_0=-0.55,j_0-\Omega_k=1$, corresponding to a ``concordance cosmology'' following flat $\Lambda$CDM with $\Omega _M=0.3$. We account for uncertainty in cosmological parameters by including a systematic error term for the expected change in the luminosity distance based on $\sigma_{q_0}=0.033$, the error in $q_0$ from the flat $\Lambda$CDM measurement of \citet{Scolnic2018TheSample}. We choose this as a conservative error in $q_0$, as it is larger than the uncertainty implied by Planck measurements, and encompasses the difference between the value $q_0=-0.55$ used in R16 and that predicted from Planck cosmological measurements \citep{PlanckCollaboration2016PlanckDepth,PlanckCollaboration2018PlanckParameters}. This approach ensures consistency with R16, and may be considered independent of the early universe since these constraints are derivable from high redshift supernovae. For a maximum redshift of $z=0.5$, this expansion diverges from the analytic luminosity distance under concordance cosmology by only 0.3\%, which is smaller than the impact of our systematic in $q_0$.

In this section we've described our use of a Taylor expanded distance-redshift relation under the assumption of isotropy and homogeneity as a baseline analysis. In order to address the effects of local voids on the Hubble diagram, we require the assumption of homogeneity to be relaxed. While a possible approach would examine the behavior of perturbations on a background FLRW metric to derive a relation between the size of a density perturbation $\delta$ and the change in the local value of the Hubble constant, we use an isotropic but inhomogeneous metric to investigate the full nonlinear impact of a spherical local void on the measured luminosity distance-redshift relation.


\subsection{Void cosmology} \label{sec:heterogcosmo}

\begin{figure}[!ht]
\epsscale{0.6}
\plotone{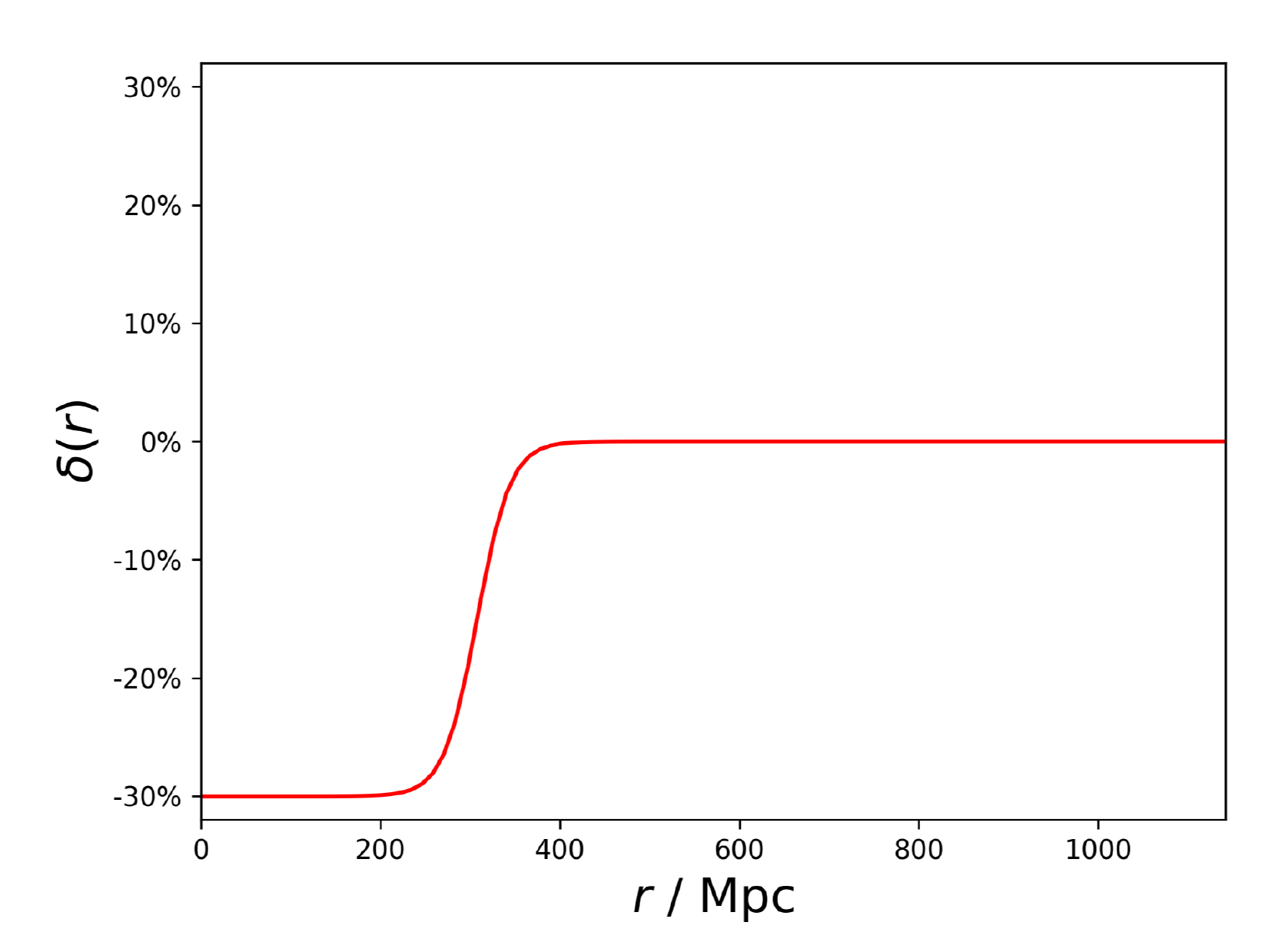}
\epsscale{0.6}
\plotone{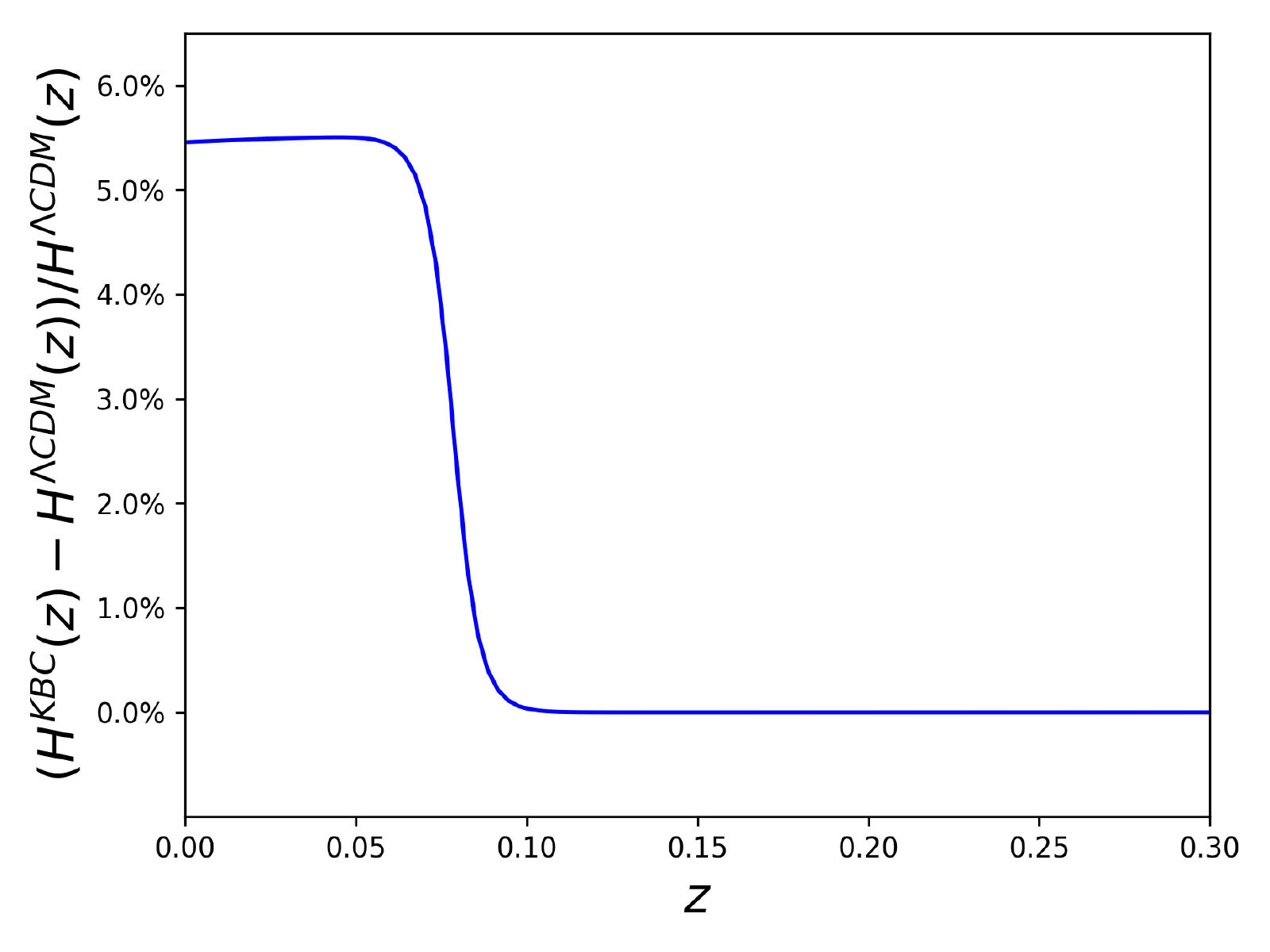}
\caption{Upper panel: Density contrast $\delta (r)$ of the KBC model as a function of radial coordinate $r$. Lower panel: Predicted expansion rate of the same KBC model as a function of $z$ relative to expansion rate of a $\Lambda$CDM cosmology with $H_0=\SI{73.2}{\kilo\meter\per\second\per\mega\parsec},\Omega_M=0.3$. The expansion rate $H(z)$ is defined as $H^{KBC}(z)=\dot{R}(r(z),t(z))/R(r(z),t(z))$ for the KBC model and $H^{\Lambda CDM}(z)=\dot{a}(z)/a(z)$ for the $\Lambda$CDM model.  \label{fig:deltaHKBC}}
\end{figure} 

To model the luminosity distance-redshift relation in the case of a local void {we follow the procedure of \citet{Hoscheit2018TheModel}. However, we argue that their boundary conditions are inappropriate, and replace them with more physically justified conditions. The model is a solution to the field equations of GR which takes the form of} the Lemaitre-Tolman-Bondi (LTB) metric \citep{Lemaitre1997TheUniverse,Tolman1934EffectModels.,Bondi1947SphericallyRelativity}, which generalizes the FLRW metric to allow for isotropic but inhomogeneous cosmologies,

\begin{equation}
ds^2=c^2 dt^2-\frac{R'(r,t)^2}{1-k(r)}dr^2-R^2(r,t)d\Omega^2
\end{equation}

where $R(r,t)$ is a generalized scale factor and $k(r)$ is a generalized curvature term, with $R'(r,t)=\partial R(r,t) / \partial r$. This metric in the homogeneous case becomes $R(r,t)=a(t)r$, $k(r)=kr^2$ and the FLRW metric is recovered. Our choice of gauge is that $R(r,0)=R_0(r)=r$. To model the mass distribution of the void we use a three parameter function { based on that of \citet{Garcia-Bellido2008ConfrontingCosmology},}

\begin{equation}
\delta(r)=\delta_V\frac{1-\tanh((r-r_V)/2\Delta_r)}{1+\tanh(r_V/2\Delta_r)} \label{eq:KBCmatterdensitycontrast}
\end{equation}

with void depth $\delta_V$, void radius $r_V$, and transition width $\Delta_r$. Modified versions of the Friedmann equations (Equations \ref{eq:sumrule},\ref{eq:friedmanneq}) with spatially dependent, spherically averaged critical densities of matter and dark energy $\Omega_M(r),\Omega_\Lambda(r)$ are recovered from the Einstein field equations,

\begin{equation}
\Omega_k(r)= \frac{-k(r) c^2}{H_0(r)^2 R_0(r)^2}
\end{equation}

\begin{equation}
1 = \Omega_M(r) + \Omega_k(r) +\Omega_\Lambda(r)  \label{eq:sumrule}
\end{equation}

\begin{equation}
\frac{\dot{R}(r,t)^2}{R(r,t)^2}=H_0(r)^2 \cdot (\Omega_M(r) \frac{R_0(r)^3}{R(r,t)^3} + \Omega_k(r)  \frac{R_0(r)^2}{R(r,t)^2} +\Omega_\Lambda(r) ) \label{eq:friedmanneq}.
\end{equation}

 Equation \ref{eq:friedmanneq} may be integrated then inverted to solve for the scale factor $R(r,t)$. The equations for null geodesics along this metric (Equations \ref{eq:timegeodesic}, \ref{eq:redshiftgeodesic}) may be numerically integrated to find cosmic time $t$ and the radial coordinate $r$ as a function of redshift, giving the redshift-luminosity distance relation through Equation \ref{eq:lumdisteq}, as described in \citet{Hoscheit2018TheModel}.

\begin{equation}
\frac{dt}{dr}=-\frac{1}{c} \frac{R'(r,t)}{\sqrt[]{1-k(r)}} \label{eq:timegeodesic}
\end{equation}
\begin{equation}
\frac{1}{1+z}\frac{dz}{dr}=\frac{1}{c} \frac{\dot{R}'(r,t)}{\sqrt[]{1-k(r)}} \label{eq:redshiftgeodesic}
\end{equation}

\begin{equation}
D^{\text{LTB}}_L(z)=(1+z)^2 R(r(z),t(z)) \label{eq:lumdisteq}.
\end{equation}

 To solve these equations, boundary terms are required to describe the spatial dependence of the critical densities $\Omega_i(r)$, as well as the Hubble constant as a function of space $H_0(r)$ \citep{Enqvist2007TheObservations}. Notably, as all cosmological parameters can vary with radial coordinate under a LTB formulation, the time since the Big Bang is allowed to vary as a function of radial coordinate subject to boundary conditions. Such a boundary condition on the time since the Big Bang $t_B(r)$ is equivalent to a condition on $H_0(r)$. Our boundary conditions are that the matter density follows the form given in \ref{eq:KBCmatterdensitycontrast}, that the energy density of dark energy ($\rho_\Lambda$) is constant with respect to space and time, that the universe is flat outside the void, and that the time since the Big Bang $t_B$ does not depend on distance from the Milky Way, respectively,

\begin{equation}
\rho_M(r) \propto \Omega_M(r) H_0(r)^2 = \Omega_{M,\text{out}}(1+\delta(r)) H_{0,\text{out}}^2  \label{eq:massdensity} 
\end{equation}
\begin{equation}
\rho_\Lambda(r) \propto \Omega_\Lambda(r) H_0(r)^2 = (1-\Omega_{M,\text{out}}) H_{0,\text{out}}^2= \text{const}  \label{eq:constlambdadensity}
\end{equation}
\begin{equation}
t_B(r) = t_B=\text{const} \label{eq:constbangtime}.
\end{equation}

Following \citet{Hoscheit2018TheModel}, we fix four of these model parameters from measurements of the KBC void and concordance cosmology at large scales; the critical matter density of the universe outside the void $\Omega_{M,out}=0.3$ and the void parameters $\delta_V=-0.3,\ r_V=\SI{308}{\mega\parsec},\ \Delta_V=\SI{18.46}{\mega\parsec}$. The Hubble constant outside the void $H_{0,\text{out}}$ (and thus $t_B$) is left as a free parameter. These assumptions then represent a universe which follows the behavior of concordance cosmology on scales $ \gtrapprox \SI{1000}{\mega\parsec}$, has minimal curvature at larger scales, and is consistent with structure formation shortly after the Big Bang. Our model with the KBC parameters predicts a change in the local value of $H_0$ of 5.5\% as seen in Figure \ref{fig:deltaHKBC}. This is different than the 3.5\% change predicted by \citet{Hoscheit2018TheModel} due to differences in the boundary conditions they used. Notably, the boundary conditions of \citet{Hoscheit2018TheModel} imply a Milky Way older than the rest of the universe by \SI{0.5}{\giga\year}; see Appendix \ref{sec:boundaryconditons} for more details. 


\section{Cosmic Voids and the Hubble constant} \label{sec:void}

\subsection{Constructing a Hubble diagram} \label{subsec:hubblediag}

\begin{figure}[!ht]
\plotone{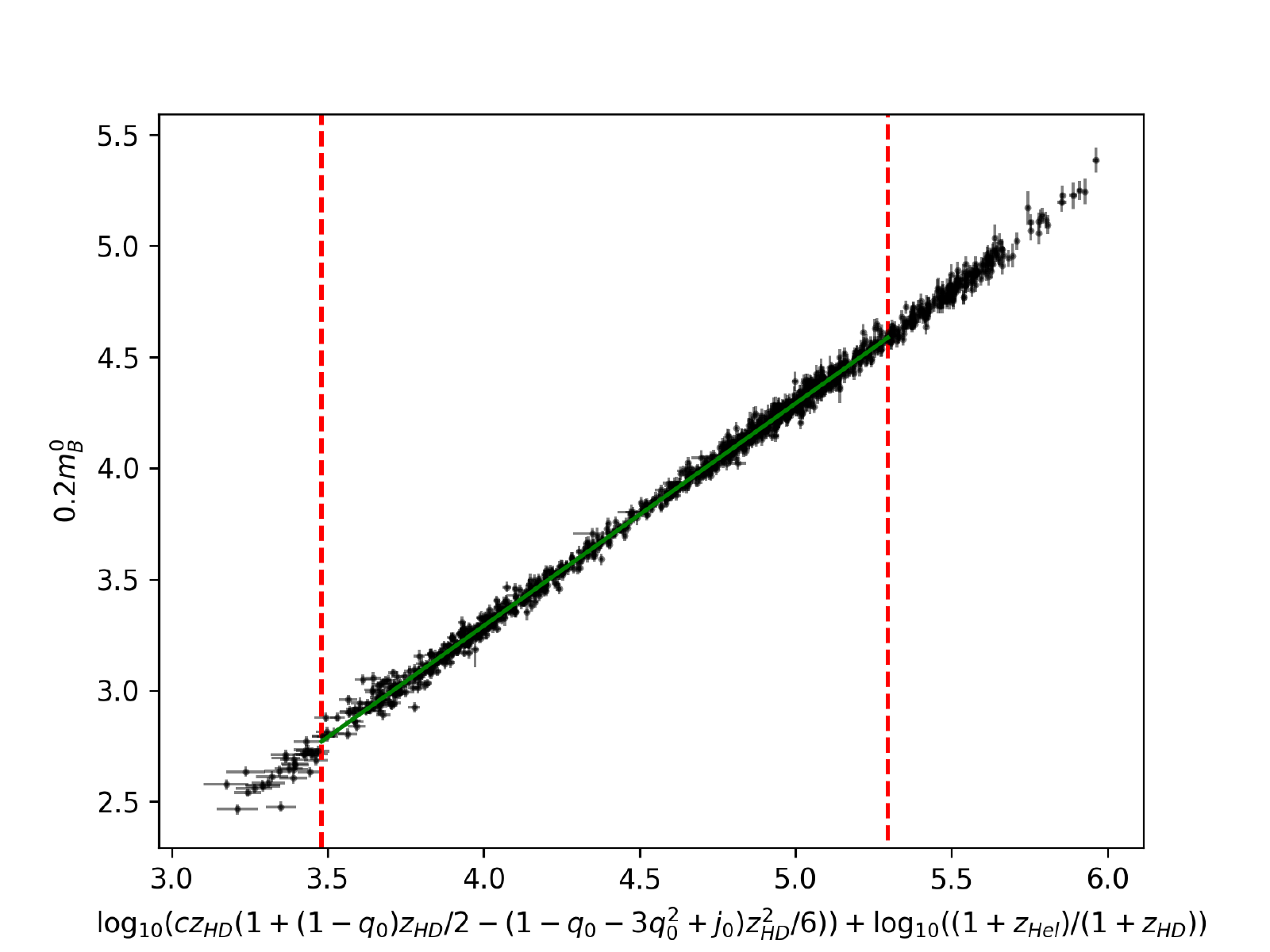}
\caption{Magnitude-redshift Hubble diagram. Intercept of the Hubble diagram determines the Hubble constant. Red lines show the edges of redshift range for primary fit, $0.01<z<0.50$
\label{fig:hubbleDiagram}}
\end{figure} 

For a supernova with corrected peak magnitude $m_B^0$ and with $M_B^0$ as the absolute magnitude of a fiducial SN Ia, the distance modulus is defined

\begin{equation}
\mu=m_B^0-M_B^0=5\log_{10}(\frac{D_L(z)}{Mpc})+25.
\end{equation}

As mentioned previously, we subtract modeled coherent bulk velocities from our low-$z$ ($z<0.08$) sample based on 2M++ mass reconstructions; however, a factor of $1+z$ in the definition of luminosity distance (see Equation \ref{eq:lumdisteq}) is due entirely to the loss of photon energy caused by redshift, and should be calculated using the \textit{apparent} redshift rather than a redshift corrected to the cosmological rest frame. This factor of $1+z$ should use the heliocentric redshift $z_{Hel}$ rather than the redshift after corrections $z_{HD}$. We account for this by inserting a term in our distance modulus

\begin{equation}
\mu=m_B^0-M_B^0=5\log_{10}(\frac{D_L(z_{HD})}{Mpc})+5\log_{10}(\frac{1+z_{Hel}}{1+z_{HD}})+25 \label{eq:distancemodulus}.
\end{equation}

This $(1+z_{Hel})/(1+z_{HD})$ term would average to no effect for a sample uniformly distributed about the sky, and has a net relative effect of order $10^{-5}$ on the measured Hubble constant for our sample. 

We retain the definition of the Hubble intercept $a_B$ from R16 as the x-intercept ($m_B^0=0$) of a Hubble diagram plotting $0.2 m_B^0$ against a (modified) $\log_{10} cz$ term, as in Figure \ref{fig:hubbleDiagram}. To measure the Hubble constant from this quantity, standard procedure is (assuming FLRW cosmology) to use Equation \ref{eq:redshiftexpansion}, and re-express Equation \ref{eq:distancemodulus} in terms of the the Hubble intercept $a_B$:

\begin{equation}
\log_{10}(\frac{H_0}{\SI{}{\kilo \metre\per\second \per\mega\parsec}})=5 + \frac{M_0^B}{5} +a_B
\end{equation}

\begin{equation}
a_{B,i}=\log_{10}(\frac{cz_{HD,i}[1+(1-q_0)z_{HD,i}/2-(1-q_0-3q_0^2+j_0-\Omega_k)z_{HD,i}^2/6+O(z_{HD,i}^3)]}{\SI{}{\kilo \metre\per\second }})+\log_{10}(\frac{1+z_{Hel,i}}{1+z_{HD,i}})-0.2m_{B,i}^0 \label{eq:interceptEquation}.
\end{equation}

The calibration of SNe Ia to the rest of the distance ladder is then encompassed in the absolute magnitude as detailed in R16 and R18, and the term $a_B$ is only dependent on the Hubble flow as measured by the SNe Ia sample. Since we are now looking for spatial variation in the Hubble flow, we take weighted averages of $a_{B,i}$ in specific redshift bins looking for evidence of variation in $a_{B,i}$ with redshift due to outflows surrounding an isotropic void. We may also fit the KBC model to values of $a_B$ by defining

\begin{equation}
a_B^{\text{KBC}}(z;a_{B,\text{out}})= \log_{10} \frac{D_L^{\text{LTB}}(z;H_{0,\text{out}}=H_{0,\text{ref}}, \text{KBC parameters})}{ D_L^{\text{FLRW}}(z;H_0=H_{0,\text{ref}},q_0=-0.55,j_0-\Omega_k=1)}+a_{B,\text{out}}.
\end{equation}

We fix the reference value of $H_{0,ref}$ since the KBC void parameters $r_V =\SI{308}{\mega\parsec}, \Delta_r=\SI{18.46}{\mega\parsec}$ are specified using the R16 value of $H_0$. Since these parameters are derived from redshift measurements, a hypothetical fractional change in the Hubble constant would require modification of these parameters by the same amount. By setting $H_{0,\text{ref}}=\SI{73.2}{\kilo\meter\per\second\per\mega\parsec}$ we encapsulate the redshift dependence of $a_B$ under the KBC model while leaving open the overall normalization of the luminosity distance (and thus the background value of $H_0$) through fitting the value of $a_B$ outside the void, $a_{B,\text{out}}$. For fits made in this section, we define the $\chi^2$ of a given fit with residuals $\boldsymbol{\delta a_B}$ as 

\begin{equation}
\chi^2= 5\boldsymbol{\delta a_B}^{\,T} \cdot \boldsymbol{C}^{-1} \cdot 5 \boldsymbol{\delta a_B} .
\end{equation}

\subsection{Effect of local void on Hubble intercept} \label{subsec:interceptAnalysis}

\begin{deluxetable*}{CcCCCC}
\tablecaption{}
\tablewidth{0pt}
\tablehead{
\colhead{Redshift range} & \colhead{Field}  & \colhead{\# of unique SNe}  & \colhead{$\Delta a_B^{z_\text{void}=0.07}$}  & \colhead{$\Delta a_B^{z_\text{void}=0.05}$} & \colhead{Change in $\chi^2$ of KBC model} 
\label{table:interceptchanges}
}
\decimalcolnumbers
\startdata
0.023<z<0.15 & Whole sky & 397 & 0.0013 \pm 0.0040 & 0.0010 \pm 0.0036 & +13.5\\
0.01<z<0.50  & Whole sky &  1054 & 0.0006 \pm 0.0036 & 0.0002 \pm 0.0034 & +26.9\\
0.01<z<0.50  & Whole sky (stat. only) &  1054 & -0.0002 \pm 0.0020 & 0.0000 \pm 0.0020 & +128.251 \\
0.01<z<0.50  & KBC Fields & 575 & -0.0031 \pm 0.0043 & - & -\\
0.01<z<0.50  & WS14 Fields & 396 & - & 0.0040 \pm 0.0045 & - \\
0.01<z<0.50  & 6dFGS-SGC &  248 & - & -0.0052 \pm 0.0064 & -  \\
\enddata
\caption{Fitted values of $\Delta a_B$ at KBC and WS14 redshifts using samples chosen with several combinations of redshift cut and choice of sky field. In no sample do we see evidence of a significant step in $a_B$. $\chi^2$ difference between KBC model predicted $a_B^\text{KBC}(z)$ and constant value of $a_B$ shown when applicable. "Stat. only" refers to a fit made without using the Pantheon systematic errors (see Section \ref{sec:snsample}) 
}
\end{deluxetable*}

\begin{figure}[!ht]
\epsscale{0.7}
\plotone{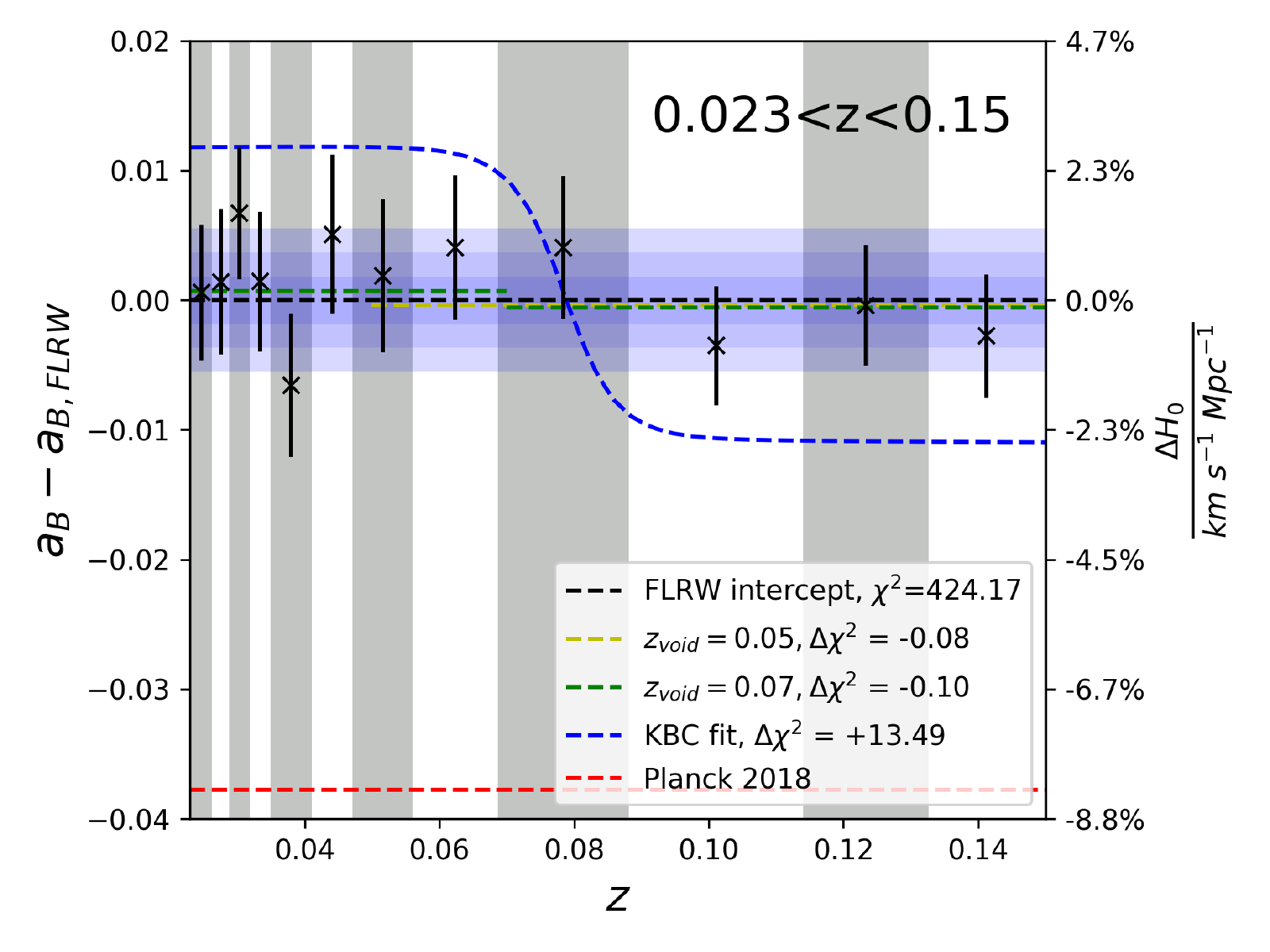}
\plotone{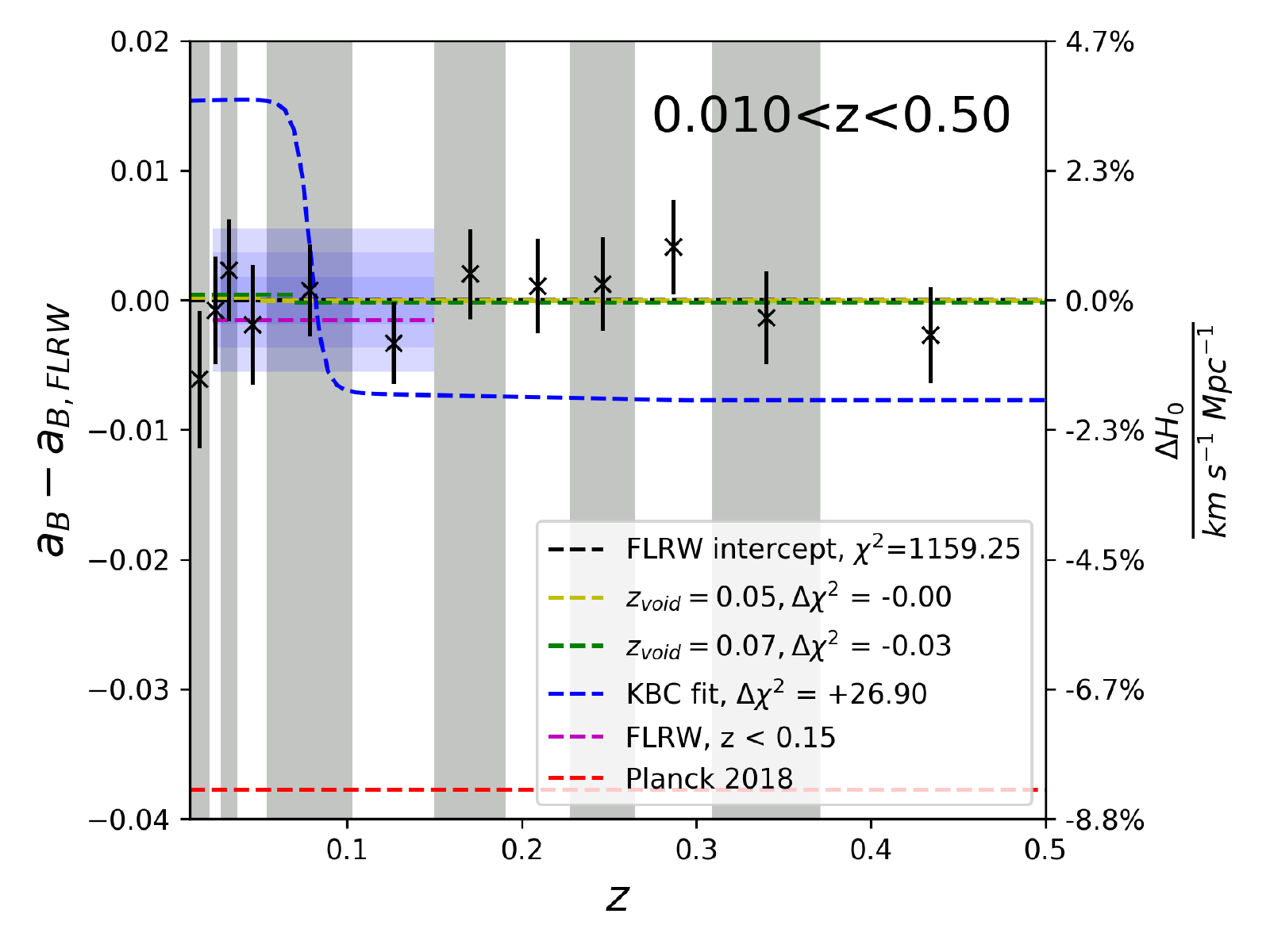}
\caption{Change in Hubble intercept from FLRW value as a function of $z$. Upper panel shows fits for 397 SNe in redshift range $0.023<z<0.15$, while lower panel shows fits for 1054 SNe with $0.01<z<0.50$. Data points are measurements of $a_B$ from supernovae binned by redshift. Dashed blue line shows best fit of $a_B^\text{KBC}(z)$ to this data. Each bin from each plot has the same number of SNe. Error bars shown only for visualization purposes, and covariances between points are not included. Regardless of analysis all intercepts are well above the value of $a_B$ required to match the Planck 2018 measurement. Blue shading shows $1\sigma,2\sigma,3\sigma$ predicted uncertainties from cosmic variance in a measurement of $H_0$ over a redshift range $0.023<z<0.15$ as found by \citet{Wu2017SampleConstant}.
\label{fig:interceptRedshiftR16Redshift}}
\end{figure} 


With the updated SNe dataset and analysis described above, our sample has expanded from the 217 used by R16 to 397 unique SNe within the R16 redshift range ($0.023<z<0.15$). For a void with a sharp edge, we can expect a step from higher $a_B$ to lower $a_B$ at the redshift corresponding to the edge of the void. We may roughly test theories of a sharp edged void by measuring the Hubble intercept in two redshift bins, $a_B^{z<z_\text{void}}$ and $a_B^{z>z_\text{void}}$, and finding the significance of a difference between them.  We find that $\Delta a_B^{z_\text{void}}= a_B^{z<z_\text{void}}-a_B^{z>z_\text{void}}$ at $z_\text{void}=0.07$ is $0.0013 \pm 0.0040$, with a significance of $0.34\sigma$, much smaller than the value $\Delta a_B^{z_\text{void}=0.07}= 0.0076$ claimed at a significance of $2.1\sigma$ by \citet{Hoscheit2018TheModel}  using the smaller R16 sample. \citet{Whitbourn2014TheRedshifts} (WS14) suggested the existence of a void within $z=0.05$. Measuring the difference in intercepts at this redshift is also insignificant with $\Delta a_B^{z_\text{void}=0.05}= 0.0010 \pm 0.0036$. These changes in $a_B$ can be converted into fractional changes in $H_0$ by multiplying by $\ln(10)$. The largest cause of the loss of significance of steps in the SN Hubble diagram is the inclusion of the Foundation survey with 141 SNe (mostly below $z=0.07$). 34 CSP supernovae that failed quality cuts using the data from \citet{Stritzinger2011THESUPERNOVAE} pass with the re-calibrated photometry from \citet{Krisciunas2017TheExplosions}, along with 21 new SNe in this redshift range. Using this model increases $\chi^2$ of the fit by 13.6 relative to an FLRW cosmology, showing that the KBC model is disfavored at $3.7\sigma$.  We show in Figure \ref{fig:interceptRedshiftR16Redshift} the best fit of the KBC model, an FLRW cosmology, and two models with two binned intercepts, one roughly corresponding to the KBC void and one roughly corresponding to the  WS14 void. None of these models shows a significant improvement over the FLRW cosmology.

The analysis of SNe Ia distances in \citet{Hoscheit2018TheModel} uses SNe within a redshift range $0.023<z<0.15$, following R16. By imposing a maximum redshift for the analysis, the R16 measurement of Hubble flow supernovae avoids excessive model-dependence from the higher order terms in the redshift expansion (see Equation \ref{eq:redshiftexpansion}). Since the statistical error in the Hubble intercept is a sub-dominant component of the error budget in $H_0$, as seen in R16, the decrease in sample size is inconsequential. The minimum redshift cut was set by \citet{Jha2007ImprovedMLCS2k2}, which found evidence in the Hubble diagram for a void at $z<0.023$ using a sample of 133 SNe. However, decreasing the redshift threshold to test for the presence of a local void is appropriate. Further, by increasing the maximum redshift threshold, we can put much stronger constraints on any cosmic void by increasing the precision in the Hubble intercept beyond a hypothetical void. Our best test for the presence of a local void uses a redshift range $0.01<z<0.5$, increasing the sample to 1054 unique SNe.

When we extend the analysis using a redshift range of $0.01<z<0.5$, as can be seen in Figure \ref{fig:interceptRedshiftR16Redshift}, the increase in statistics gives a substantially increased constraint on the Hubble intercept at high redshift $a_B^{z>z_\text{void}}$, and thus $\Delta a_B^{z_\text{void}}$.  The step in intercept at $z=0.07$ loses further significance becoming $\Delta a_B^{z_\text{void}=0.07
}=0.0006 \pm 0.0035$ ($0.17\sigma$). This is $6.2\sigma$ discrepant with the $\Delta a_B^{z_\text{void}=0.07}=0.0232$  predicted from the expected 5.5\% change in $H_0$ for an instantaneous transition to a $\delta=-30\%$ void. An FLRW cosmology fit to this data has $\chi^2=1159.25$ ($\nicefrac{\chi^2}{DoF}=1.07$). Fitting the KBC model increases this $\chi^2$ by 28.6, corresponding to a $5.3\sigma$ rejection of the model. From the figure, it is clear that there are no substantial negative residuals relative to an FLRW cosmology at high redshift. We conclude that the KBC model is inconsistent with the SN data at $>5\sigma$. Similarly, the isotropic void of \citet{Shanks2019iGaia/iTension}, with $\delta=-20\%$ to a redshift of $z=0.05$ predicts a value of $H_0$ about 3.5\% larger within this redshift, corresponding to $\Delta a_B^{z_\text{void}=0.05}=0.0153$. The measured value of a step in $a_B$ at $z=0.05$ is $\Delta a_B^{z_\text{void}=0.05
}= 0.0002 \pm 0.0034$, and is inconsistent with this prediction at $4.5\sigma$. The measured values of $\Delta a_B$ are shown in Table \ref{table:interceptchanges}.

We note that neither of these models is capable of moving the R18 value of $H_0$ by the 8.3\% necessary to resolve the tension between distance-ladder measurements and $\Lambda$CDM predictions of $H_0$ based on early-universe measurements, as can be seen from the line on plots above showing the value of $a_B$ required to resolve the Hubble tension. Both are inconsistent with the SN Hubble diagram at $>4\sigma$. We conclude that the SN Hubble diagram disfavors local voids described by these models.

\subsection{Void Isotropy}

\begin{figure}
\plotone{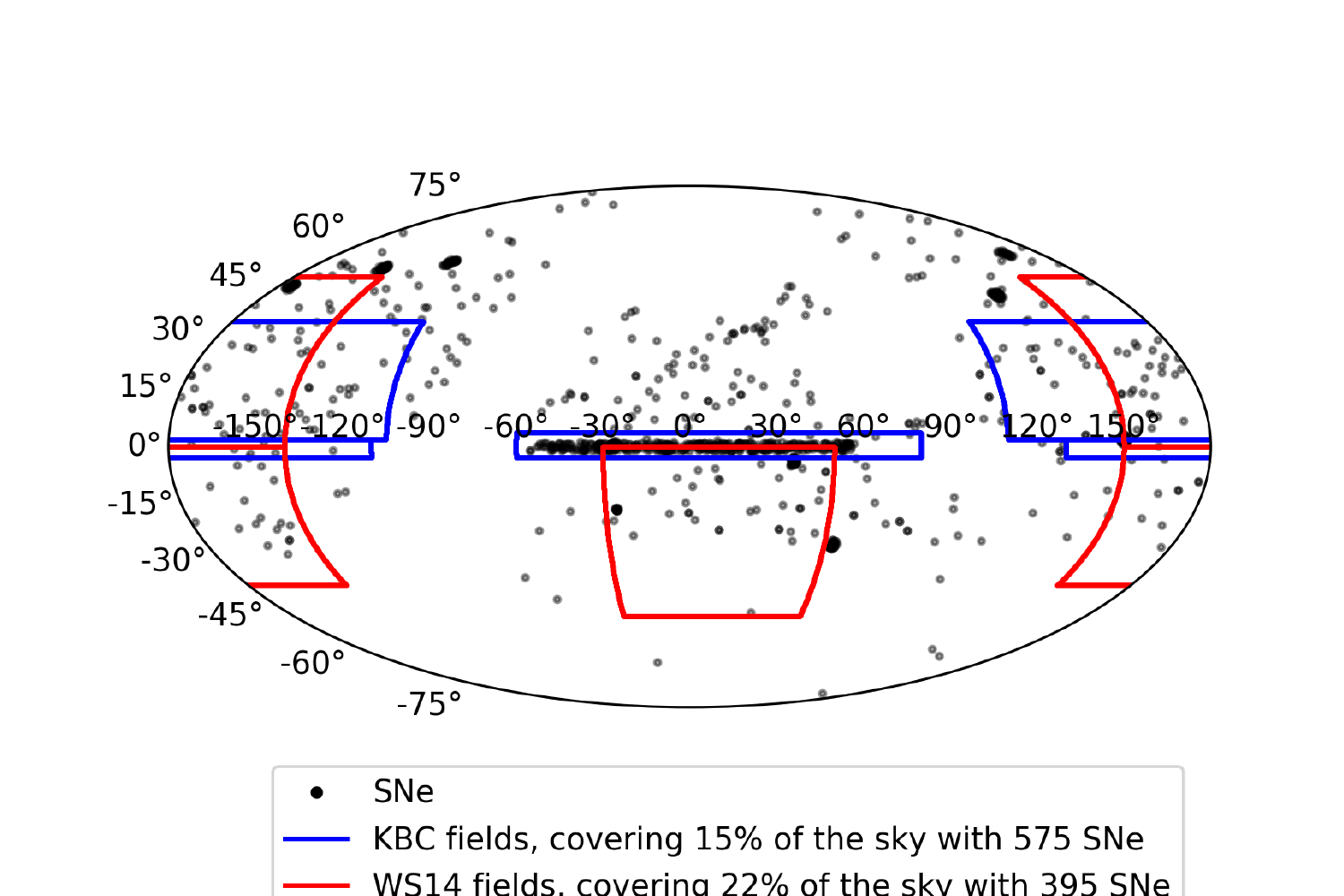}
\caption{ Distribution of SNe Ia across the sky using Mollweide projection. The fields encompassing the data used by KBC are shown in red, the fields used by WS14 are shown in blue, and the SNe are shown in black.
\label{fig:skypositions}}
\end{figure}

\begin{figure}
\epsscale{0.7}
\plotone{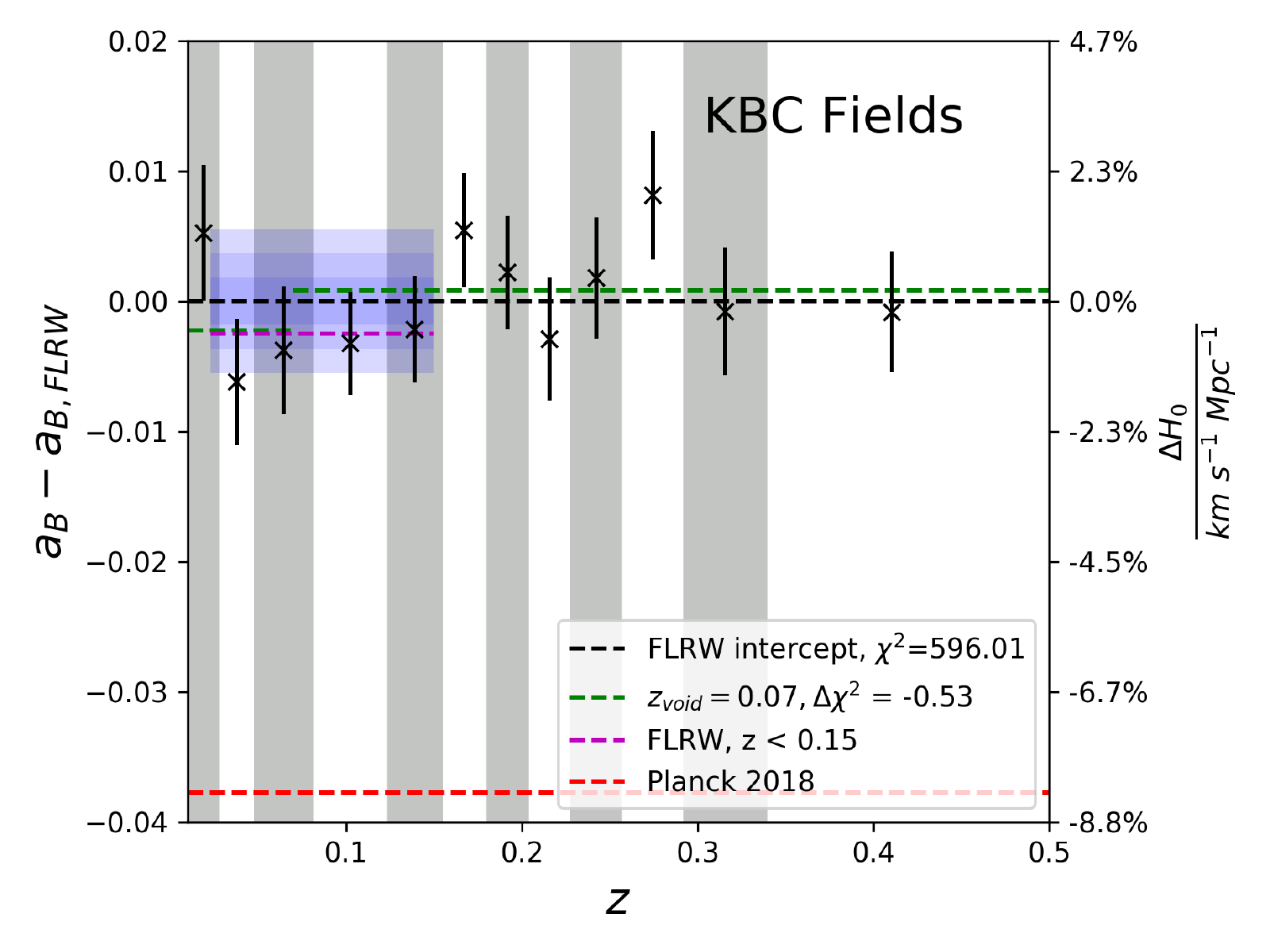}
\plotone{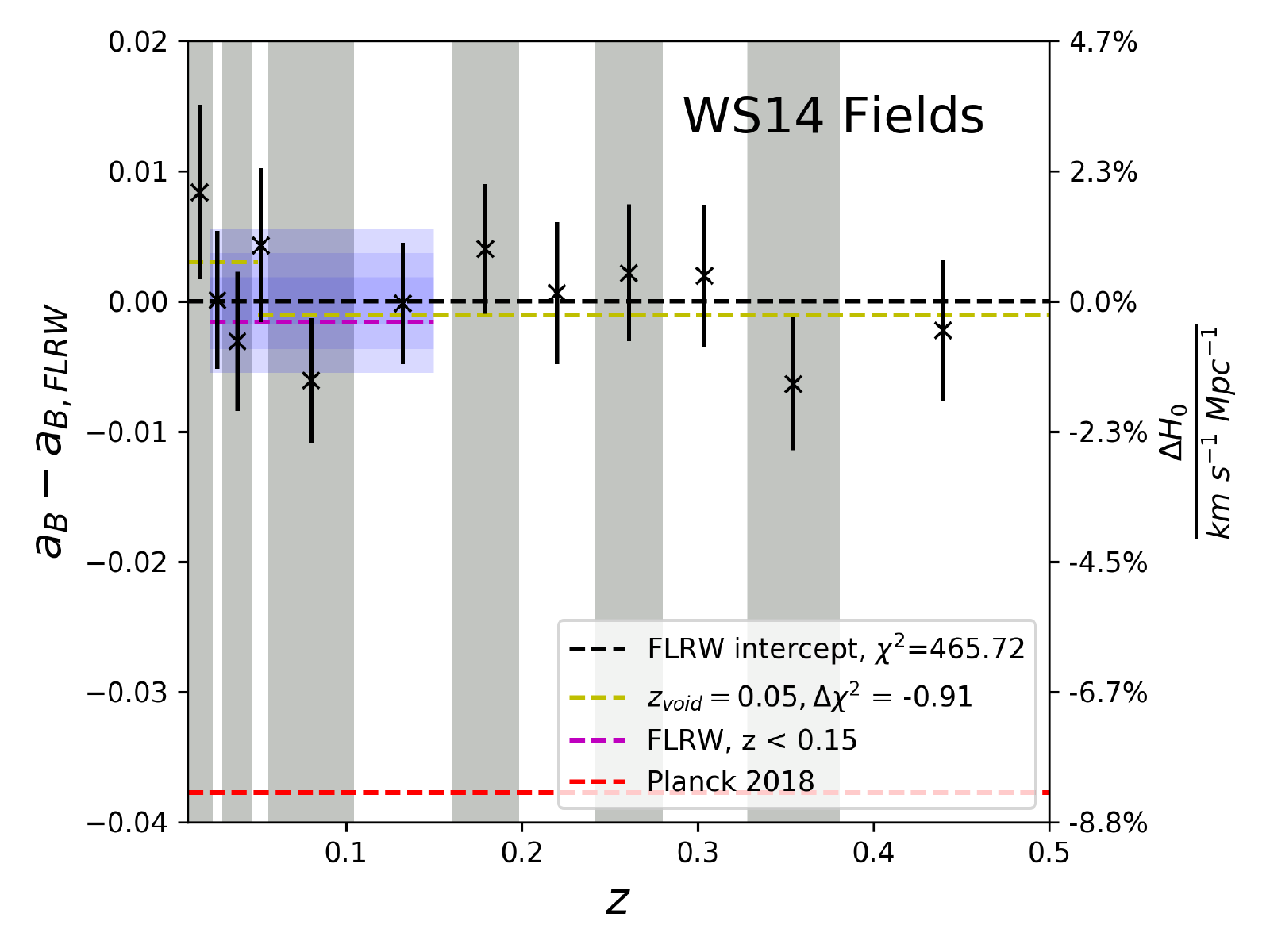}
\caption{Upper panel: Change in Hubble intercept from FLRW value as a function of $z$ from $z=0.01$ to $z=0.50$ for supernovae contained within the KBC fields. Lower panel: Change in Hubble intercept from FLRW value as a function of $z$ from $z=0.01$ to $z=0.50$ for supernovae contained within the WS14 fields. Data points are binned intercept values. Each bin in each plot has the same number of SNe.  Error bars shown only for visualization purposes, and covariances between points are not included. Blue shading shows $1\sigma,2\sigma,3\sigma$ predicted uncertainties from cosmic variance in a measurement of $H_0$ over a redshift range $0.023<z<0.15$ as found by \citet{Wu2017SampleConstant}.
\label{fig:kbcfields}}
\end{figure}
Both models of local density contrasts we have discussed so far extrapolate results measured only from a minority of the sky to the entire sky. The galaxy catalogue used by \citet{Keenan2013EVIDENCEDISTRIBUTION} is estimated to reliably cover a total of $\SI{584}{\square {deg}}$ (1\% of the sky) using the ``counts-in-cell'' method . When investigating the isotropy of their measured data, they compare results using galaxies from three broad areas on the sky which encompass their individual patches of high estimated sample completeness, shown in Figure \ref{fig:skypositions}, which cover in total $\SI{6172}{\square {deg}}$ (15\% of the sky). Similarly, \citet{Whitbourn2014TheRedshifts} drew data from three fields totaling $\SI{9162}{\square {deg}}$ (22\% of the sky). The SNe Ia distribution across the sky is extremely inhomogeneous, since surveys such as SDSS repeatedly returned only to certain fields. It is therefore plausible that measured under-densities are anisotropic, explaining why measurements of the Hubble intercept that assume isotropy (after 2M++ bulk flow corrections) and use an inhomogeneous data set do not register the deep voids seen by these studies. While evidence from the CMB dipole disfavors many off-center void models, certain peculiar velocities of the Milky Way  would allow more extreme void geometries \citep{Enqvist2007TheObservations}. As seen in Figure \ref{fig:skypositions}, the  distribution of observed SNe Ia over the sky overlaps to some extent with these fields, allowing us to look for evidence of outflows in these directions by repeating the analysis with supernovae contained entirely within these fields. In general, analytic prediction of the effects of anisotropic and inhomogeneous cosmologies is extremely difficult. We thus look for outflows by examining the data for steps in the intercept $a_B$ in these directions without fitting the KBC model. Since we are now looking for anisotropic effects (possibly including those in 2MASS++), we remove the \citet{Carrick2015CosmologicalField} flow corrections to redshift from our data to leave any possible features in place. 

Restricting our analysis to the SNe contained within the KBC fields reduces the number of unique SNe in the dataset to 575. Nevertheless, the analysis done with only SNe from the KBC fields shows no evidence of outflows. The measured size of a step in $a_B$ at $z=0.07$ decreases to $\Delta a_B^{z_\text{void}=0.07}=-0.0031 \pm 0.0043$. As a result, there is no evidence from the SNe Ia sample of an outflow in the directions constrained by \citet{Keenan2013EVIDENCEDISTRIBUTION}, and this dataset is in conflict with the predicted size of an outflow stemming from a the Hubble diagram at $5.6\sigma$. There are fewer SNe in the fields from WS14, totaling 396 unique SNe. The difference between intercepts above and below $z=0.05$ is $\Delta a_B^{z_\text{void}=0.05} = 0.0040 \pm 0.0045$, a result $2.6\sigma$ away from the predicted 3.5\% change in $H_0$ from \citet{Shanks2019iGaia/iTension}.

While KBC found no evidence of density contrast variation between their three fields, WS14 found different density contrasts in the three fields they studied, up to a redshift of $z=0.05$. In the 6dFGS-NGC and SDSS-NGC fields, we do not have enough SNe to constrain outflows relative to the predictions of WS14 in these directions. However the 6dFGS-SGC field has much better coverage, and we look for a step in $a_B$ at $z=0.05$ corresponding to their measurements. For the 6dFGS-SGC field, with 248 unique SNe, the change in $a_B$ at $z=0.05$ is $\Delta a_B^{z_\text{void}=0.05}= -0.0052 \pm 0.0064$, and is $5.9\sigma$ discrepant with the predicted change in $H_0$ from the WS14 density contrast in this direction of $\delta=-40\%$. 

We  conclude that we can exclude at $>5\sigma$ the effects of proposed large ($>25\%$) anisotropic density contrasts on the Hubble diagram, and disfavor the effects of smaller voids.

\subsection{General void cosmologies} \label{subsec:voiddepth}
\begin{figure}[!ht]
\plotone{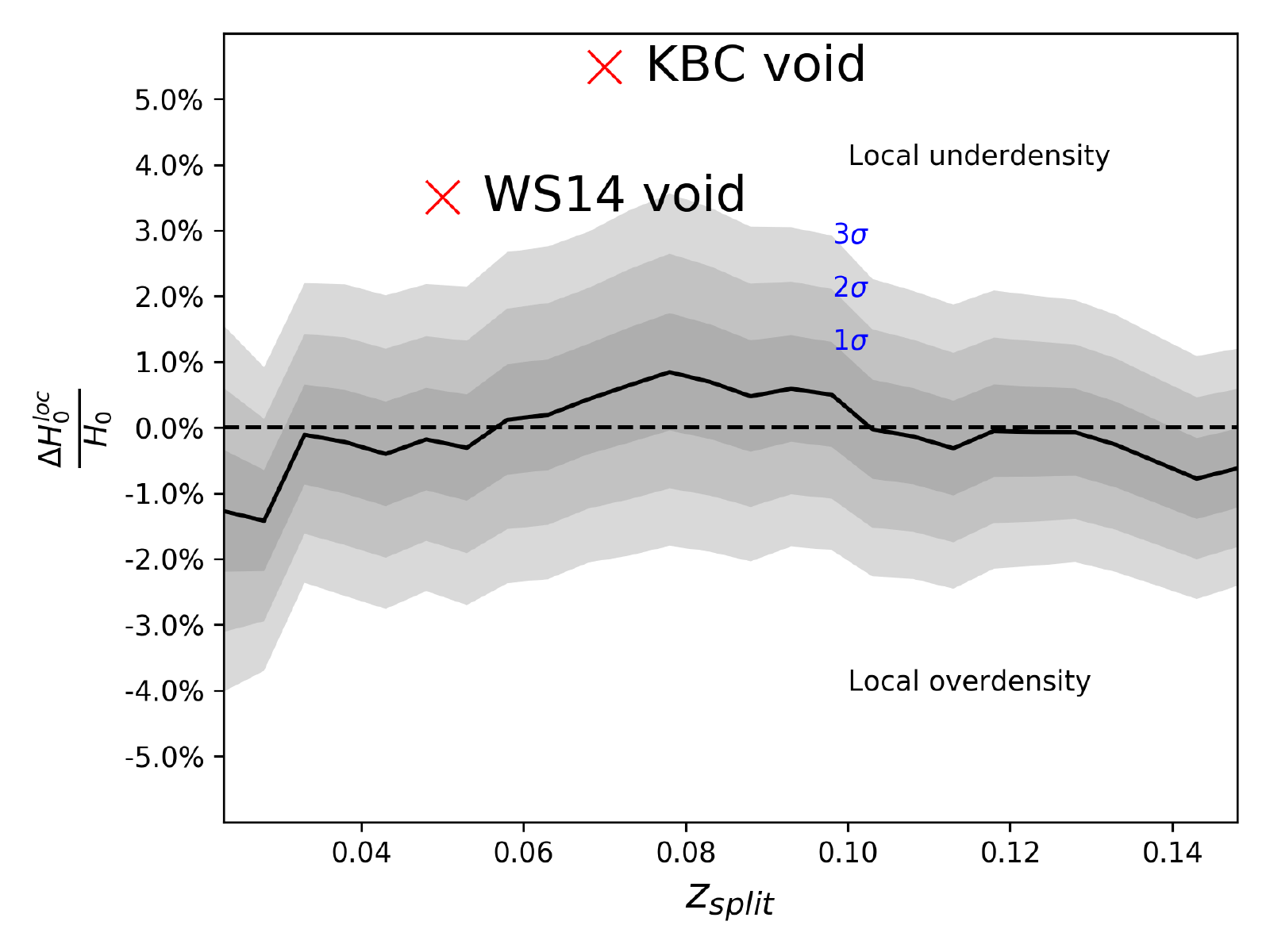}
\caption{Difference between values of $H_0$ measured above and below $z_\text{split}$ using SNe in redshift range $0.01<z<0.5$. $z_\text{split}$ is allowed to vary over the R16 redshift range. Red crosses show expected change in $H_0$ for KBC and WS14 voids.
\label{fig:voidDepth}}
\end{figure}

To this point, we have mostly focused on specific examples of outflows. Nevertheless, the SN Ia data shows little evidence of a cosmic void \textit{at any redshift}. To illustrate this, in Figure \ref{fig:voidDepth} we divide the SNe into two redshift bins, and allow the boundary $z_\text{split}$ between redshift bins to vary. Looking for the largest step in intercept value from this plot will correspond to the worst possible interpretation of our data. The largest possible significance in this sample is still $<2\sigma$ (at $z_\text{split}=0.023$). 
Since we find that no sectioning produces a difference in Hubble intercepts greater than $2\sigma$ within the R16 redshift range, this is a strong indicator that any effects of local matter density contrasts on our measurement are small. Under this analysis, further, we see no sign of the $z<0.023$ void reported in \citet{Jha2007ImprovedMLCS2k2}. This is consistent with the results of \citet{Conley2007IsGalaxies}, which found no evidence of a void after addressing issues with light-curve modelling and fitting.

Due to systematic effects such as the transition from the low-$z$ sample to SDSS at $z \approx 0.08$ and statistical effects of sample size, our ability to constrain void models is strongly dependent on their physical size. Our constraint is loosest at $z=0.023$ with an uncertainty of 0.94\% in $H_0$, and strongest at $z=0.15$ with an uncertainty of 0.60\% in $H_0$. Given that at these redshifts we see no evidence of a significant step in $a_B$, it is appropriate to state that we see no variation in $H_0$ as a function of $z$ at redshifts $z>0.023$ with a $1\sigma$ precision of $0.94\%$, corresponding to a $5\sigma$ constraint on local density contrasts on scales larger than $\SI{69}{\mega\parsec}\ h^{-1}$ of $|\delta| <  27\%$.

To quantify the total effect of any local structure on the R18 measurement, we may measure the significance of any difference between the intercept measured over our the high end of our redshift range $0.15<z<0.5$ and the intercept measured over the R16 redshift range $0.023<z<0.15$. $a_B^{0.023<z<0.15}-a_B^{0.15<z<0.5}=-0.0023 \pm 0.0026$, corresponding to a precision in $H_0$ of 0.60\%. Based on \citet{Wu2017SampleConstant}, we would expect $H_0^{0.023<z<0.15}$ to have cosmic variance of approximately 0.42\%, or approximately 0.0018 for $a_B^{0.023<z<0.15}$. Our result is then consistent with the predicted cosmic variance from \citet{Wu2017SampleConstant}, since we see no significant variation and our empirical errors are larger than this effect. 

\section{Discussion and Conclusions} \label{sec:conclusion}

Our work looks for evidence of outflows in the SNe Ia Hubble diagram that would impact the determination of $H_0$. We create a sample of distance and redshift measurements of cosmological SNe by combining data from the Pantheon sample with the Foundation survey and the most recent release of lightcurves from the Carnegie Supernova Project. We conclude that the distance-redshift relation of this sample is inconsistent with the large local void proposed by \citet{Keenan2013EVIDENCEDISTRIBUTION} at $5.3\sigma$, that of \citet{Shanks2019iGaia/iTension} at $4.5\sigma$, and find no evidence of a change in the Hubble constant corresponding to a void with a sharp edge at any redshift used in \citet{Riess2016ACONSTANT}. From our analysis we derive a $5\sigma$ constraint on local density contrasts on scales larger than $\SI{69}{\mega\parsec}\ h^{-1}$($z<0.023$) of $|\delta| <  27\%$.  

In comparison with the work of \citet{Hoscheit2018TheModel} and \citet{Shanks2019iGaia/iTension}, which found marginal evidence for the effect of local voids upon the Hubble diagram, our study uses a larger sample of low-redshift SNe than either. Further, neither of these studies accounted for systematic uncertainties in the SNe data, uncertainties which we have estimated based on the analysis of \citet{Scolnic2018TheSample}. Neglecting these systematics leads to the artificially low errors in $\Omega_M$ seen in \citet{Shanks2019iGaia/iTension} of $\sigma_{\Omega_M}=0.01$ (compared to $\sigma_{\Omega_M}=0.022$ from \citet{Scolnic2018TheSample} ) as well as overestimation of the significance of the results of both analyses.  The effects of systematics on our analysis are significant, as seen in Table \ref{table:interceptchanges}, contributing $\approx 70\%$ of the variance in our primary results. 

A reconciliation of the results of \citet{Keenan2013EVIDENCEDISTRIBUTION} or \citet{Whitbourn2014TheRedshifts} with this study would require that there be unquantified systematic uncertainties. \citet{Scolnic2018TheSample} specifically budgets for 85 known systematics in the SN data. While we have not repeated this analysis for the new SNe included in our sample, we have set the size of these systematics to be equal to past surveys. Furthermore our analysis uses the same scatter model and nuisance parameters as \citet{Riess2016ACONSTANT}. Bias corrections used in the full Pantheon analysis averaged over the G10 scatter model (used in this work and R16) and the C11 model \citep{Chotard2011TheWidths}, which has increased chromatic variation between SNe relative to the G10 model. Nonetheless, our use of a single scatter model and choice of parameters are appropriate for comparison to relevant prior work such as R16, \citet{Hoscheit2018TheModel}, and \citet{Wu2017SampleConstant}. Further, our systematic error budget includes the expected effect from the uncertainty in these fit parameters on each SN, and a systematic of half the difference between biases from the G10 and C11 scatter models, which will reduce (although not eliminate) the effects of these choices. Similarly, we have minimized dependence on high-redshift cosmology through inclusion of an error in $q_0$ and a redshift cut $z<0.5$. Given the value of $\sigma_{q_0}$ we have chosen, cosmological model revisions could reduce the significance at which we disfavor local void models, but such a revision would still fail to move the distance ladder measurement by enough to resolve the Hubble tension, and simultaneously introduce further tensions with Planck measurements of $\Omega_M$, $\Omega_k$.

There are many systematic uncertainties in galaxy luminosity function measurements, some of which are unquantified. The fits of \citet{Keenan2013EVIDENCEDISTRIBUTION} have much better $\chi^2$ in some redshift bins than in others, which could be an indication of model failure or of anisotropies in the underlying data. These problems could be exacerbated by issues with spectroscopic completeness, spectroscopic selection biases, source confusion, or scale dependence of the linear bias parameter. That \citet{Whitbourn2014TheRedshifts} measures density contrasts between -40\% and -5\% in different fields perhaps suggests that errors in these measurements are larger than stated in their work. Indeed, \citet{Shanks2019iGaia/iTension} implicitly adopts this position by assuming that these measurements reflect an underlying isotropic void of density contrast -20\%. 

Reducing the dependence of SN Ia distance-ladder measurements on local structure will require a better understanding of the local universe, as well as the reduction of systematics in low-$z$ SNe through better calibrated cosmological SNe from surveys like Foundation \citep{Foley2018TheRelease,Jones2018ShouldEnvironments}. Another possible route is to increase the minimum redshift of the sample to avoid dependence on local structure. This will be made easier as increased numbers of SNe at higher redshifts are made available from surveys such as LSST \citep{Ivezic2008LSST:Products,TheLSSTDarkEnergyScienceCollaboration2018TheDocument}. While we cannot generally exclude all possible effects of local structure on SN measurements, our work provides a strong constraint on simple examples thereof and confirms the suitability of distance-ladder methods for measurement of the Hubble constant to a precision approaching $1\%$.

\acknowledgements
We would like to thank David Jones at UC Santa Cruz for his assistance in calculating host galaxy stellar masses used in this work.

\appendix
\section{Boundary conditions of Lemaitre-Tolman-Bondi model} \label{sec:boundaryconditons}
Our boundary conditions in Equations \ref{eq:massdensity}, \ref{eq:constlambdadensity}, and \ref{eq:constbangtime} are different from those of \citet{Hoscheit2018TheModel} which neglects the fact that $H_0(r)$ is radially dependent within a LTB framework, and thus that the critical densities $\Omega_i(r)$ and physical densities $\rho_i(r)$ do not have the same radial dependence. Their boundary condition uses $\delta(r)$ as the change in the \textit{critical} density of matter $\Omega_M(r)$, while our boundary condition self-consistently uses $\delta(r)$ as the change in the \textit{physical} density $\rho_M(r)$. This distinction leads to the absence of the factors of $H_0(r)^2$ in their boundary conditions. Furthermore \citet{Hoscheit2018TheModel} does not provide their boundary condition on $H_0(r)$ or $t_B(r)$. We infer from their results that their boundary condition was similar to $H_0(r) \propto t_B(r)$, in contrast with our Equation \ref{eq:constbangtime}. Using the KBC void parameters, these changes lead us to predict a change in $H_0$ from inside to outside the void $\approx 60\%$ larger than the prediction of \citet{Hoscheit2018TheModel}. We consider the choices of boundary condition of \citet{Hoscheit2018TheModel} unmotivated, as their boundary conditions imply spatial variation in the physical density of dark energy and that the region around the Milky Way is older than the rest of the universe by $ \approx \SI{0.5}{\giga\year}$. Our predictions match those of \citet{Marra2013CosmicParameter}. While \citet{Hoscheit2018TheModel} claim their prediction is consistent with that of \citet{Wu2017SampleConstant}, \citet{Wu2017SampleConstant} does not show a relation between the change in $H_0$ from background within a putative void and density contrast of that void comparable to these predictions; their Figure 4 shows a simulated relation between change in $H_0$ when measured at $z<0.15$ and the density contrast at $z<0.04$. We conclude that our boundary conditions are more appropriate for a model with the goal of matching the results of concordance cosmology at large scales while modifying the local value of the Hubble constant.

%
%

\bibliographystyle{aasjournal}
\bibliography{Mendeley}


\end{document}